\documentclass[aps,pre,amsmath,amssymb,lengthcheck,showpacs,superscriptaddress]{revtex4-1}
\usepackage[dvipdfmx]{graphicx}
\usepackage{color}
\usepackage{amsfonts}
\usepackage{physics}
\usepackage{braket}
\usepackage{multibib}
\usepackage[english]{babel}
\usepackage{bm}

\begin{document}
\title{Phase transition in the binary mixture of jammed particles with large size dispersity}
\author{Yusuke Hara}
\email{hara-yusuke729@g.ecc.u-tokyo.ac.jp}
\affiliation{Graduate school of arts and science, University of Tokyo, Komaba, Tokyo 153-8902, Japan}
\author{Hideyuki Mizuno}
\affiliation{Graduate school of arts and science, University of Tokyo, Komaba, Tokyo 153-8902, Japan}
\author{Atsuhi Ikeda}
\affiliation{Graduate school of arts and science, University of Tokyo, Komaba, Tokyo 153-8902, Japan}
\affiliation{Research Center for Complex Systems Biology, Universal Biology Institute, University of Tokyo, Komaba, Tokyo 153-8902, Japan}

\date{\today}

\begin{abstract}
It has been well established that particulate systems show the jamming transition and critical scaling behaviors associated with it.
However, our knowledge is limited to (nearly) monodisperse systems. 
Recently, a binary mixture of jammed particles with large size dispersity was studied, and it was suggested that two distinct jammed phases appeared. 
Here, we conduct a thorough numerical study on this system with a special focus on the statistics of and finite-size effects on the fraction of small particles that participate in the rigid network. 
We present strong evidence that two distinct jammed phases appear depending on the pressure and composition of two species, which are separated by the first-order phase transition.
In one of two phases, only large particles are jammed, whereas both small and large particles are jammed in the other phase.
We also describe the phase diagram in the pressure-composition plane, where the first-order phase transition line terminates at a critical point.
In addition, we investigate the mechanical properties in terms of the elastic moduli over the phase diagram and find that discontinuous changes in elastic moduli emerge across the phase transition.
Remarkably, despite the discontinuities, the elastic moduli in each jammed phase exhibit identical scaling laws to those in the monodisperse systems.
\end{abstract}


\maketitle


\section{\label{Intro}Introduction}
Jammed particulate systems are ubiquitous in our lives.
Emulsions, colloidal suspensions, and granular materials are examples of jammed systems, where constituent particles are randomly jammed~\cite{VanHecke2010}. 
The random structure of jammed systems is one source of difficulties in understanding these systems. 

One of the simplest models for jammed particulate systems is the assemblies of athermal particles that interact via short-range repulsive potentials.
When we compress these particles from the low density, the system gains rigidity at the density called the jamming point.
This phenomenon is known as the jamming transition established by many previous works, e.g., Ref.~\cite{OHern2003}.
The geometrical, mechanical, and vibrational properties of the system are known to follow the critical power-law near the transition where the distance from the jamming points plays the role of a control parameter~\cite{OHern2003,Silbert2005,Wyart2005,Ellenbroek2009a,Wyart2010,Vitelli_2010,Zaccone2011,Goodrich2012,Goodrich2014,Lerner2014,DeGiuli2014,Karimi2015,Mizuno2017,Shimada_2018,Mizuno2018}. 
Recently, jammed systems composed of dimer-shaped particles were shown to exhibit the same critical laws as a sphere packings~\cite{Schreck_2010,Shiraishi2019,Shiraishi_2020}.
In the last two decades, numerical and theoretical studies of these systems have developed to a large extent, and we have a good level of understanding of the critical behaviors near the jamming transition in nearly monodisperse jammed particles. 

However, the jammed particles in reality are not monodisperse and are composed of particles of multiple sizes~\cite{Huang2001,Zeng2015,Kwok2020}. 
In particular, when the sizes of larger and smaller particles are quite different, these systems can exhibit phenomena that are never observed in the monodisperse systems~\cite{Derkach2009,Clara-Rahola2015}.
For example, mixtures of large colloidal particles and small polymers are known to show rich phenomena, including the emergence of two distinctive colloidal glass phases and a peculiar nonlinear mechanical response~\cite{Poon2002,Sciortino2005,Sentjabrskaja2013,Sentjabrskaja2019}.

Despite its theoretical and practical importance, the impacts of the polydispersity with large size ratio on the characteristic features of jammed packings are poorly understood.
The simplest system to study the effects of polydispersity can be a binary mixture of particles with large size dispersity.
The packings of binary mixtures have been studied to improve the packing efficiency.
It is now well known that an injection of fine particles to coarse large structures realizes the random close packing with a higher packing fraction than monodispersity~\cite{YERAZUNIS1965,Zheng1995,Farr2009,Meng2014,Koeze2016, Prasad2017,Pillitteri2019}.

In addition, the nature of the jamming transition in binary systems was studied. 
Xu \textit{et al.} showed that binary mixtures of equal numbers of small and large particles did not change the critical behavior near the jamming transition even if their sizes were significantly different~\cite{Xu2010}.
Kumar \textit{et al.} studied the effects on the bulk moduli by injecting a few percent of fine particles into a volume of large particles~\cite{Kumar2016}.
More recently, it was suggested that this system exhibited two different types of jammed phases, and there was a transition between them~\cite{Prasad2017,2020arXiv200202206P}.
One phase is characterized by the jamming of only large components, and the other is characterized by the jamming of both components. 
Two different types of jammed states are also observed for two dimensional discs~\cite{Koeze2016}.

In the present work, we further investigate the structural and mechanical properties of the packing of binary mixtures of particles with large size dispersity. 
In particular, we thoroughly study the transition between two jammed phases, which was very recently reported in Refs.~\cite{Prasad2017,2020arXiv200202206P}. 
We focus on the fraction of small particles that contribute to the rigidity of the system as the order parameter and study their statistics and finite-size effect. 
We show strong evidence that there is a phase transition between two phases and that the transition is first-order in nature. 
We also reveal the phase diagram of this system and find that the first-order transition line terminates at a critical point. 
In addition, we show that the mechanical properties exhibit a discontinuous change at the first-order transition. 

This paper is organized as follows.
In Sec.~\ref{Method}, we introduce the detailed description of the system in interest.
In Sec.~\ref{sec:results}, we present our simulation results.
In Sec.~\ref{sec:transition}, we focus on the analysis of the structural characterization at the fixed pressure and establish the first-order transition. 
Then, in Sec.~\ref{sec:phase-diagram}, we extend the analysis to a broad range of pressure and reveal the phase diagram. 
Finally, in Sec.~\ref{sec:mech}, the mechanical properties such as the bulk and shear moduli are studied.
We provide the summary and conclusion in Sec.~\ref{conclusion}. 

\section{\label{Method}Numerical method}
\subsection{System description}
We study the jammed packings composed of the binary mixtures of large and small particles in three spatial dimensions. 
The diameter of the large particle is $6$ times that of the small particles. 
We denote the volumes of large and small particles as $v_l$ and $v_s$, respectively, where $v_l = 6^3 v_s$. 
The numbers of large and small particles are denoted by $N_L$ and $N_S$, respectively.
These particles interact via a finite-range, purely repulsive potential~\cite{OHern2003}; two particles interact with each other only when they are in contact.
The interaction potential is the following harmonic potential: 
\begin{equation}
v_{ij}(r_{ij}) = \frac{\epsilon}{2} \left(r_{ij} - \sigma_{ij} \right)^{2} \Theta \left(1 - \frac{r_{ij}}{\sigma_{ij}} \right)
\end{equation}
where $\sigma_{ij}$ is the sum of the radii of particles $i$ and $j$, $r_{ij}$ is the distance between the centers of particles $i$ and $j$, and $\Theta(x)$ is Heaviside step function; $\Theta(x) = 1$ for $x>0$, otherwise $\Theta(x) = 0$.
We set the unit of the length and the energy as the diameter of the small particle and $\epsilon$, respectively. 

The state point of this system can be characterized by two relevant parameters. 
The first one is the volume ratio of small particles: 
\begin{equation}
X_S=\frac{N_S v_s}{N_S v_s + N_L v_l}. 
\end{equation}
In this work, we focus on the range of $0.05 \leq X_S \leq 0.25$, where the number of small particles is approximately $10$ to $70$ times that of large particles.
The second is the pressure:
\begin{eqnarray}
P=-\sum_{\left<ij \right>} r_{ij} \frac{dv_{ij}(r_{ij})}{dr_{ij}},
\end{eqnarray}
where $\sum_{\left<ij \right>}$ is the summation over all contacting pairs of particles.
The packing fraction is sometimes used as the control parameter, but we use the pressure because it enables us to more properly study the critical behaviors. 
Packings with desired $(P,X_S)$ are obtained by iterations of compression and decompression of the system, as described below.

There are three possible quantities to define the system size: $N_S+N_L$, $N_S$, and $N_L$. 
Although previous studies~\cite{Prasad2017,2020arXiv200202206P} of binary mixtures use the total number of particles $N_S+N_L$ to express the system size, we can also use the number of small particles $N_S$ or large particles $N_L$. 
In this work, we use $N_S$ to precisely study the size effect on the susceptibility ~(see Sec.~\ref{sec:transition}), while we use $N_L$ to explore the broad range of $X_S$~(see Sec.~\ref{sec:phase-diagram}).

\subsection{\label{PG}System preparation}
We generate the jammed packing of desired pressures using the iterative compression and decompression of the system and the FIRE algorithm to minimize the energy~\cite{PhysRevLett.97.170201}.
The condition to terminate the relaxation algorithm is $ \max_{i, \alpha} {\left( F_i^{\alpha} \right)}\leq 10^{-13}$, where Greek index $\alpha$ is the spatial index, and Roman index $i$ is the index of the particles.

We prepare the configuration below the jamming point by relaxing the random configuration with a low packing fraction to the mechanical equilibrium.
Then, we compress the system with fixed increment $\Delta \phi_{ini}=0.01$, and the systems are relaxed to mechanical equilibrium in each step.
We continue to increase the packing fraction until the pressure of the equilibrated configuration exceeds the target pressures.
The procedures are terminated if the pressure of the configuration is consistent with the target pressures within the acceptable error; in this case, $\delta P=|P-P_{tag}| \leq10^{-2} P_{tag}$.
When this condition is not satisfied, we (1) decompress with the rate $\Delta \phi$ and (2) compress with the new rate $\Delta \phi_{new}=0.5 \Delta \phi$.
The iterative use of the procedures brings the system to the desired pressure.

We generate at least $100$ packings at each state point. 
We denote the average of physical quantities $x$ in the ensemble of the packings generated at a state point by $\langle x \rangle_{\rm sample}$.
We also analyze the statistics of quantities $x$ in the ensemble and calculate the probability distribution denoted as $P(x)$.
Note that the ordering of large particles occurs in some packings.
We detect these packings by calculating the local bond order parameter of the structure of large particles~\cite{Lechner2008}.
These packings are rare, and we exclude them from our statistical ensemble.

\section{Results}~\label{sec:results}
\subsection{\label{sec:transition}Transition between two jammed phases}

\begin{figure}[t]
\includegraphics[width=9.0cm]{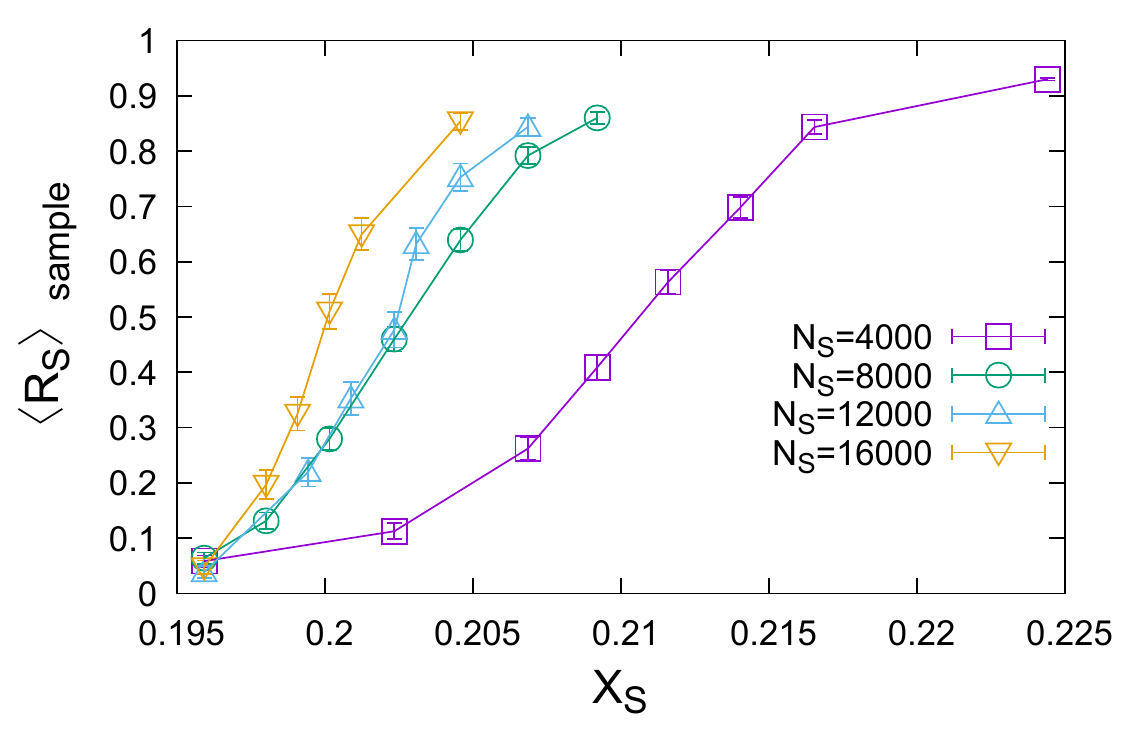}
\caption{
Mean value of the order parameter $\langle R_S \rangle_{\rm sample}$ versus $X_S$ at $P=10^{-3}$. 
Different symbols correspond to system sizes $N_S=4000, 8000, 1200, 16000$. 
The number of samples is 400 for $N_S=4000, 8000$ and 200 for $N_S=12000, 16000$ respectively. 
The increase in $\langle R_S \rangle_{\rm sample}$ becomes steeper when the system size increases. 
}
\label{fig:AveRS-NS}
\end{figure}

\begin{figure}[t]
\includegraphics[width=9.0cm]{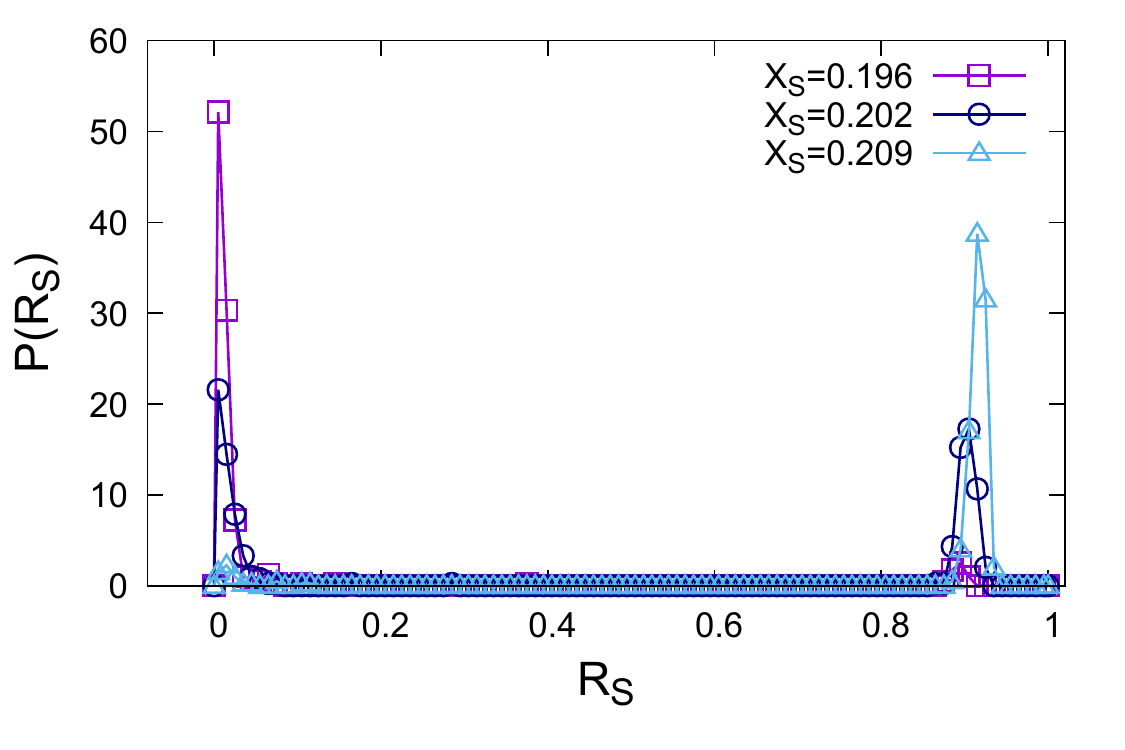}
\caption{
Probability distributions of order parameter $R_S$ over 400 independent packings at $P=10^{-3}$ for $N_S=8000$. 
The distribution becomes bimodal at $X_S = 0.202$. 
}
\label{fig:HISTRS-P-3}
\end{figure}

\begin{figure}[t]
\includegraphics[width=9.0cm]{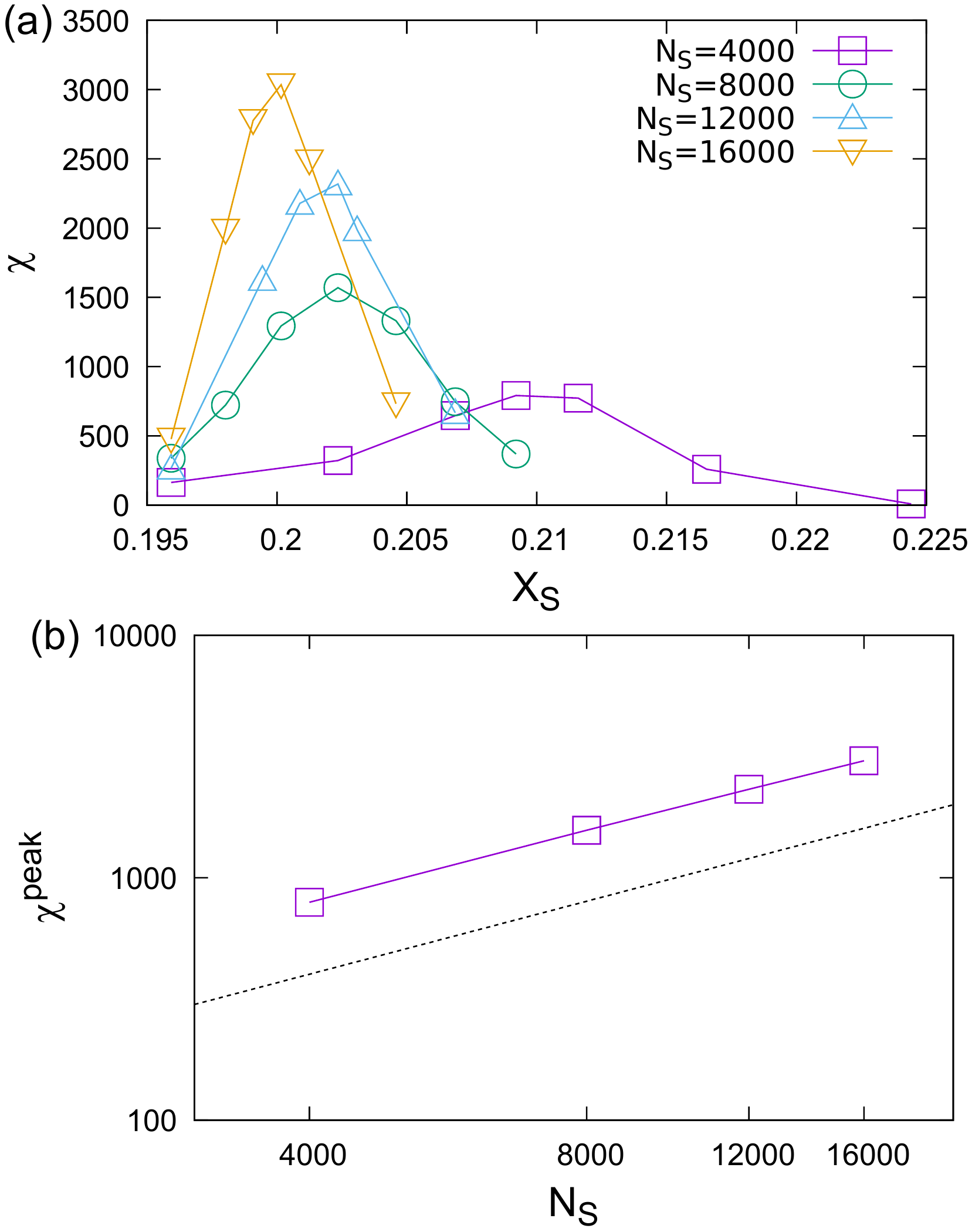}
\caption{
(a) Susceptibility (variance of the order parameters) $\chi$ versus $X_S$ at various system sizes $N_S$. 
The susceptibility exhibits the strong finite size effects. 
(b) Peak height of the susceptibility in the top panel, which is denoted by $\chi^{\rm peak}$. 
The dotted line indicates the linear dependence on system size $N_S$.
The pressure is $P=10^{-3}$.
}
\label{fig:var-size}
\end{figure}

\begin{figure}[t]
\includegraphics[width=9.0cm]{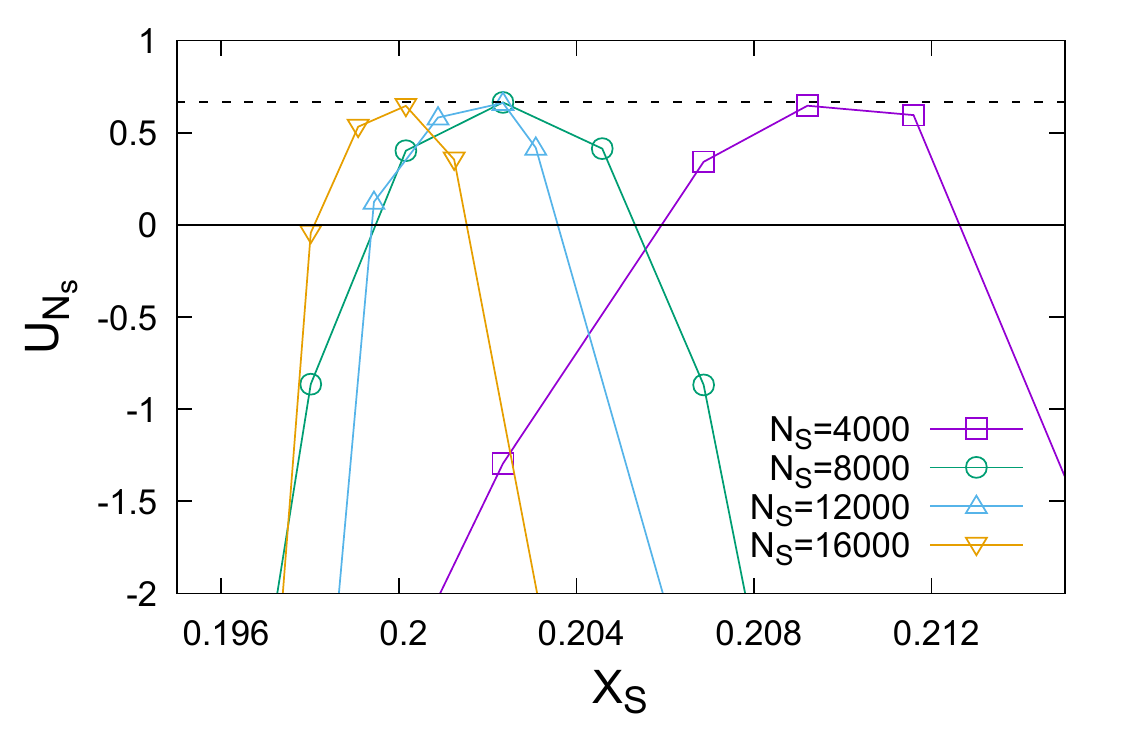}
\caption{
Binder parameter $U_{N_S}$ versus $X_S$. 
The horizontal dashed line indicates $2/3$.
}
\label{fig:binder}
\end{figure}

In this section, we study the structural properties of the packings with changing $X_S$ at the fixed pressure $P = 10^{-3}$. 
We will show that the system exhibits two distinct jammed phases, which are separated by a first-order phase transition. 
One phase is named the L phase, where only large particles are jammed, and the other is the LS phase, where both large and small particles are jammed.
To characterize these phases, we focus on the fraction of small particles that participate in the connected network of particles: 
\begin{equation}
R_S = \frac{(N_S-N_S^r)}{N_S}. 
\end{equation}
Here, $N_S^r$ is the number of small particles that are rattlers, which are defined as the particles whose contact number is less than $d+1 =4$~(where $d=3$ is spatial dimension). 
We prepare at least $200$ packings at each state point $(P=10^{-3},X_S)$ and calculate $R_s$ for each packing. 

First, we focus on $\langle R_S \rangle_{\rm sample}$, which is the mean value of $R_S$ in the ensemble of at least $200$ packings at each state point. 
Fig.~\ref{fig:AveRS-NS} shows $\langle R_S \rangle_{\rm sample}$ versus $X_s$ for various system sizes $N_S$.
$\langle R_S \rangle_{\rm sample}$ monotonically increases with $X_s$; namely, a larger fraction of small particles corresponds to more small particles participating in the connected network. 
Notably, the change in $\langle R_S \rangle_{\rm sample}$ becomes steeper when the system size increases.
This implies there is a phase transition between the states with $R_S \approx 0$ and $R_S \approx 1$ in the thermodynamic limit, and $R_S$ is an order parameter in this phase transition.

To clarify whether this is a genuine phase transition, we analyze the statistics of $R_S$. 
First, Fig.~\ref{fig:HISTRS-P-3}
 shows the probability distribution of the order parameter $R_S$ over 400 different packings for several state points. 
When $X_S = 0.196$, the distribution has a single peak at $R_s \approx 0$; most small particles are rattlers. 
When $X_S = 0.209$, the distribution also has a single peak, but it is located at $R_s \approx 1$; most small particles participate in the connected network of particles. 
At $X_S = 0.202$, the distribution shows the double peaks at $R_s \approx 0$ and 1, respectively; most small particles are rattlers in some packings, while most small particles participate in the connected network of particles in the other packings.
The appearance of the bimodal distribution strongly suggests that this is the first-order phase transition between two phases. 

Next, we consider the system size dependence of this behavior. 
To facilitate the analysis, we introduce ``susceptibility" $\chi$, which is defined as
\begin{equation}
\chi = N_S \left(\langle R_S^2 \rangle_{\rm sample} - \langle R_S \rangle_{\rm sample}^2 \right). 
\end{equation}
Fig.~\ref{fig:var-size}(a) shows $\chi$ versus $X_S$. 
Clearly, $\chi$ exhibits a peak, and the peak height increases with the system size. 
When the transition is first-order in nature, the peak height is expected to linearly depend on the system size. 
To study this point, Fig.~\ref{fig:var-size}(b) shows the system size dependence of the peak height of the susceptibility, which is denoted by $\chi^\text{peak}$. 
The data are consistent with the linear dependence on system size $N_S$, which confirms that the transition between two phases is first-order. 

Finally, we evaluate the reduced cumulant of the order parameter defined by 
\begin{equation}
U_{N_S} = 1 - \frac{\langle \left( R_S - \langle R_S \rangle_{\rm sample} \right)^4 \rangle_{\rm sample}}{3 \left( \langle \left( R_S - \langle R_S \rangle_{\rm sample} \right)^2 \rangle_{\rm sample} \right)^2},
\end{equation}
which is known as the binder parameter~\cite{Binder1984a}. 
Fig.~\ref{fig:binder} shows $U_{N_S}$ versus $X_S$ at various system sizes. 
Regardless of the system size, the binder parameter converges to $\frac{2}{3}$, which indicates the appearance of the bimodal distribution and a common feature of the first-order transition.
This plot also shows the negative dip of the binder parameter, which is another common feature of first-order transitions.

In summary, we provided strong evidences for the first-order transition in the jammed phase, which is the transition between the jammed phase with only large components jamming (L phase) and that with both components jamming (LS phase).

\subsection{\label{sec:phase-diagram}Phase diagram}

\begin{figure}[t]
\includegraphics[width=9.0cm]{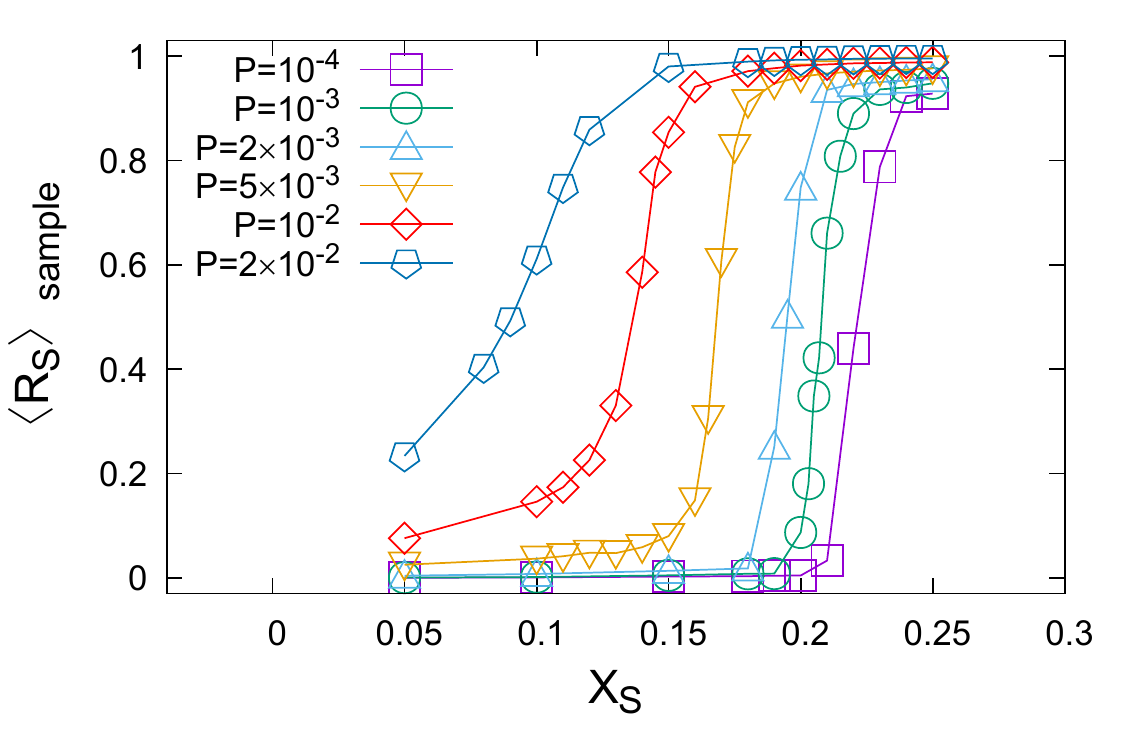}
\caption{
Mean value of order parameter $\langle R_S \rangle_{\rm sample}$ versus $X_S$ at various pressures. 
}
\label{fig:AveRS}
\end{figure}

In the previous section, we established the presence of the first-order transition at $P=10^{-3}$, where fraction $R_S$ of small particles participating in the connected network is the order parameter. 
In this section, we extend the analysis to the broad range of pressures and determine the phase diagram of the packing of a binary mixture of particles. 
As a consequence, we will show that the first-order transition line terminates at a finite pressure. 

Fig.~\ref{fig:AveRS} plots $\langle R_S \rangle_{\rm sample}$ versus $X_S$ at various pressure. 
At low pressure, $R_S$ exhibits discontinuous jumps around $X_S \approx 0.2$, which suggests the first-order transition as discussed in the previous section.
However, at high pressure, e.g., $P=2 \times 10^{-2}$, the increase becomes much milder. 
At this pressure, $R_S$ appears to smoothly change along $X_S$, which implies the disappearance of the first-order transition.

\begin{figure}[t]
\includegraphics[width=9cm]{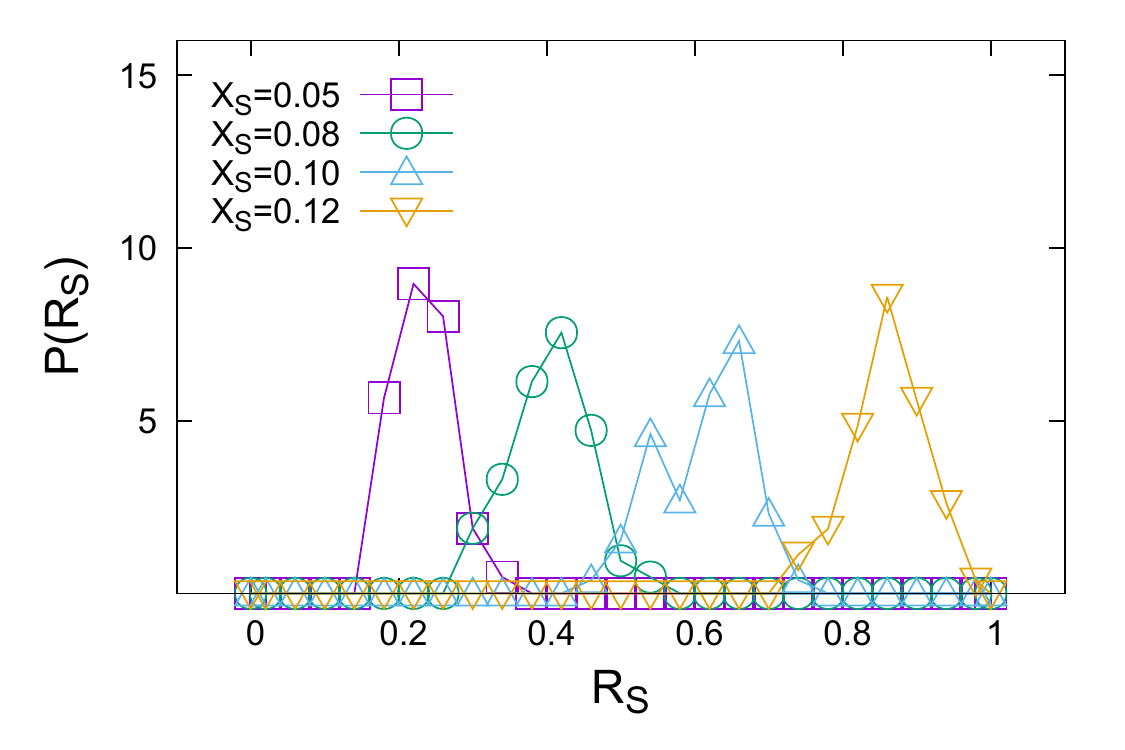}
\caption{
Probability distributions of order parameter $R_S$ at $P=2\times10^{-2}$. 
The distribution has only a single peak, and the peak continuously moves with $X_S$. 
}
\label{fig:HISTRS-P20-2}
\end{figure}

\begin{figure}[t]
\includegraphics[width=9cm]{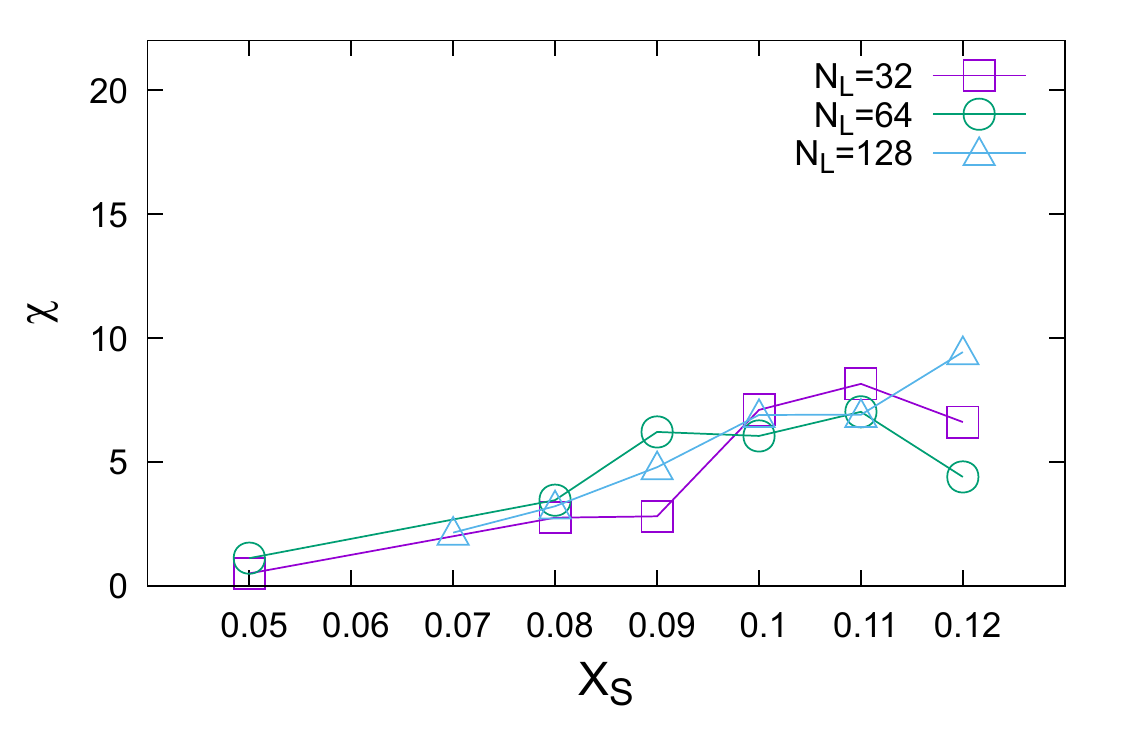}
\caption{
Susceptibility (variance of the order parameters) $\chi$ versus $X_S$ at various system sizes $N_L$.
The pressure is $P=2 \times 10^{-2}$.
The susceptibility does not show the system size dependence. 
}
\label{fig:VARRS-P20-2}
\end{figure}

To confirm this point in more detail, we analyze the statistics of $R_S$ at high pressure as performed at low pressure in the previous section. 
Fig.~\ref{fig:HISTRS-P20-2} shows the distribution of order parameter $R_s$ over different packings at $P=2\times10^{-2}$. 
The evolution of the distribution with $X_S$ is completely different from that at $P=10^{-3}$. 
The distribution always has only a single peak at all $X_S$, and the peak only continuously shifts with $X_S$. 
In addition, Fig.~\ref{fig:VARRS-P20-2} plots $\chi$ versus $X_S$ at the same pressure. 
Clearly, $\chi$ does not depend on the system size, which is a totally different observation in Fig.~\ref{fig:var-size}. 
These distinctive behaviors provide strong evidences for the lack of the first-order transition at higher pressure $P=2\times10^{-2}$. 
Thus, there is a critical point at a finite pressure $P_c$ between $P=10^{-3}$ and $P=2\times10^{-2}$. 

\begin{figure}[t]
\includegraphics[width=9.0cm]{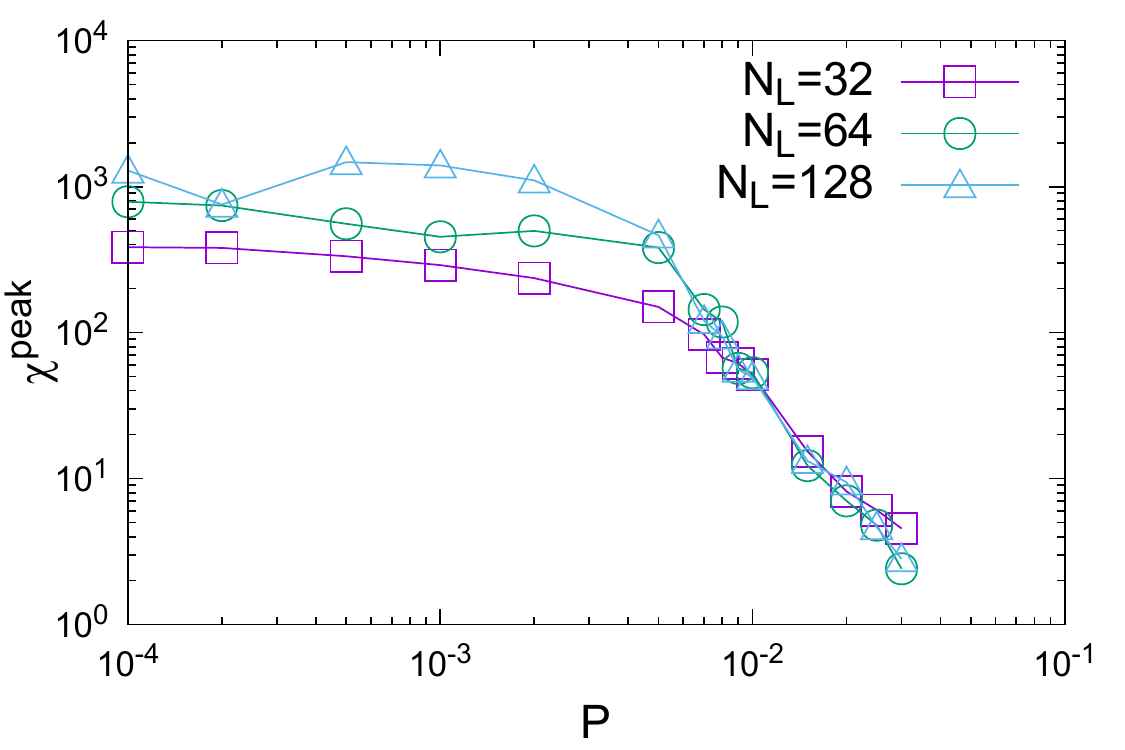}
\caption{
The peak height of susceptibility $\chi^{\rm peak}$ is plotted as a function of the pressure. 
The strong system-size dependence appears at approximately $P=2 \times 10^{-3}$.
}
\label{fig:chi-var-peak}
\end{figure}

\begin{figure}[t]
\includegraphics[width=9.0cm]{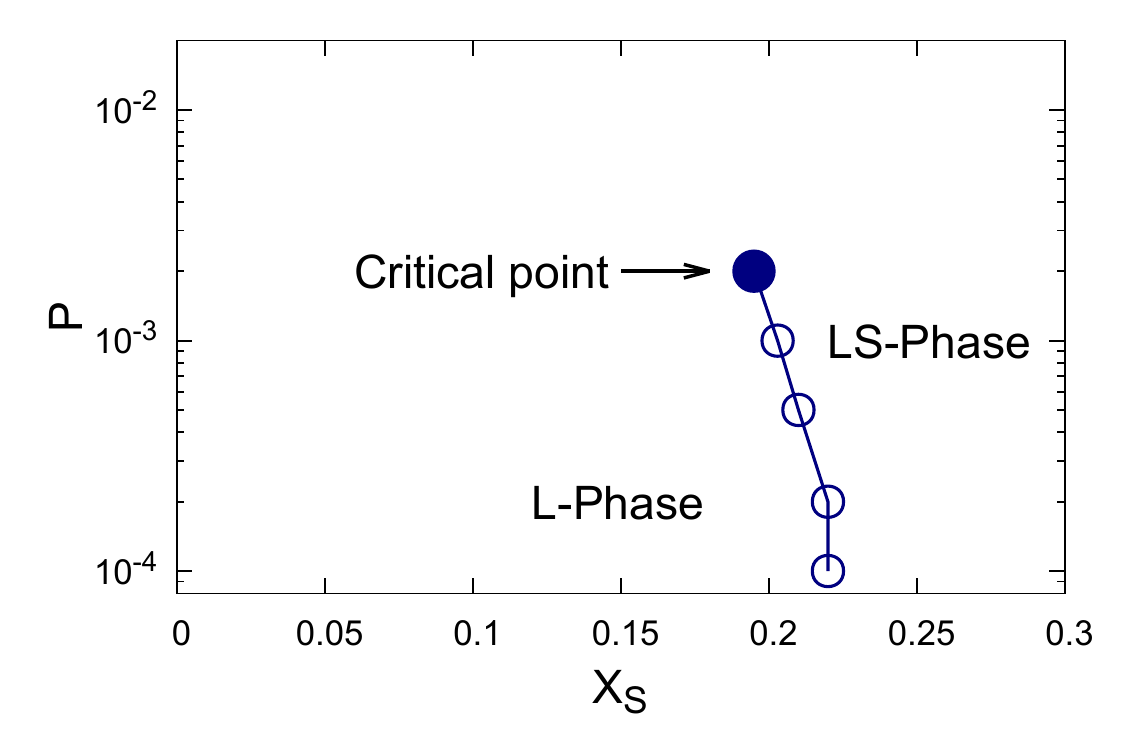}
\caption{
Phase diagram of the packings of binary mixtures of particles. 
}
\label{fig:PD}
\end{figure}

We repeated the above analysis at different pressures, plotted $\chi$ versus $X_S$ and observed the peak. 
At pressure $P$, we denote the peak height by $\chi^{\rm peak}(P)$ and its position in $X_S$-axis by $X_S^{\rm peak}(P)$. 
In Fig.~\ref{fig:chi-var-peak}, $\chi^{\rm peak}(P)$ versus $P$ is plotted for various system sizes. 
$\chi^{\rm peak}(P)$ depends on the system size at low pressure and not at high pressure. 
We determine that the first-order transition occurs at pressure $P$ if $\chi^{\rm peak}(P)$ exhibits the strong system size dependence even in our largest system size. 
Then, the first-order transition point is estimated as $(P,X_S^{\rm peak}(P))$, which are shown as open symbols in Fig. \ref{fig:PD}. 
We also observe that $\chi^{\rm peak}(P)$ increases with decreasing $P$ in the high-pressure regime, which suggests that $\chi^{\rm peak}$ diverges at critical pressure $P_c$. 
To precisely determine $P_c$, we must simulate much larger systems, which is beyond the scope of this work and can be addressed in the future. 
Here, we provide a rough estimate of $P_c$ as the highest pressure at which we observe the strong size effects in $\chi^{\rm peak}$: $P_c \approx 2 \times 10^{-3}$, which is shown as a closed symbol in Fig. \ref{fig:PD}. 
In this phase diagram, the LS phase indicates the state with $R_S \approx 1$, where both components are jammed, and the L phase is the state with $R_S \approx 0$, where only the large component is jammed.
At jamming point $P \to 0$, the first-order transition is located at $X_S = 0.22$, and the transition line continues for a finite pressure.
This transition line is terminated at the critical point, which is estimated to be at $X_S=0.195$ and $P=2 \times 10^{-3}$.

\subsection{\label{sec:mech}Mechanical properties}

\begin{figure}[t]
\includegraphics[width=\columnwidth]{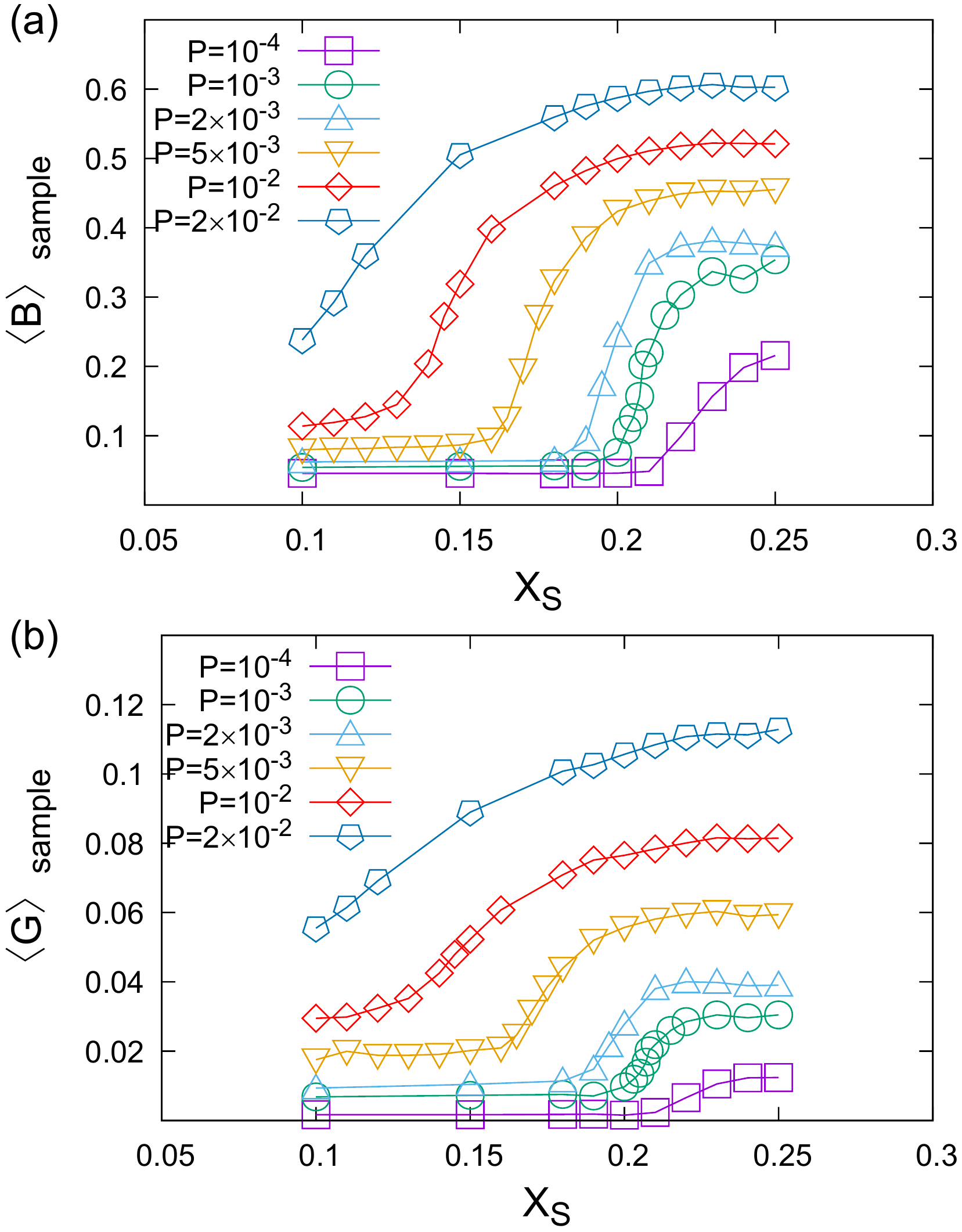}
\caption{
(a) Mean value of bulk modulus $B$ versus $X_S$. 
The system size is fixed at $N_L=64$. Each symbol represents the value at a different pressure.
(b) Same as (a) but for shear modulus $G$
}
\label{fig:Ave-mod}
\end{figure}

So far, we have established the phase diagram of the binary mixtures where there is the first-order transition line, and it terminates at the critical pressure. 
In this section, we explore the impact of this phase behavior on the mechanical properties of the system. 
We investigate bulk modulus $B$ and shear modulus $G$ of the system.
Both moduli are calculated from the linear response formalism. 
Details of this calculation and formulation are provided in Appendix \ref{CE}.
We present the results for $N_L=64$ in this section.

\begin{figure}[t]
\includegraphics[width=\columnwidth]{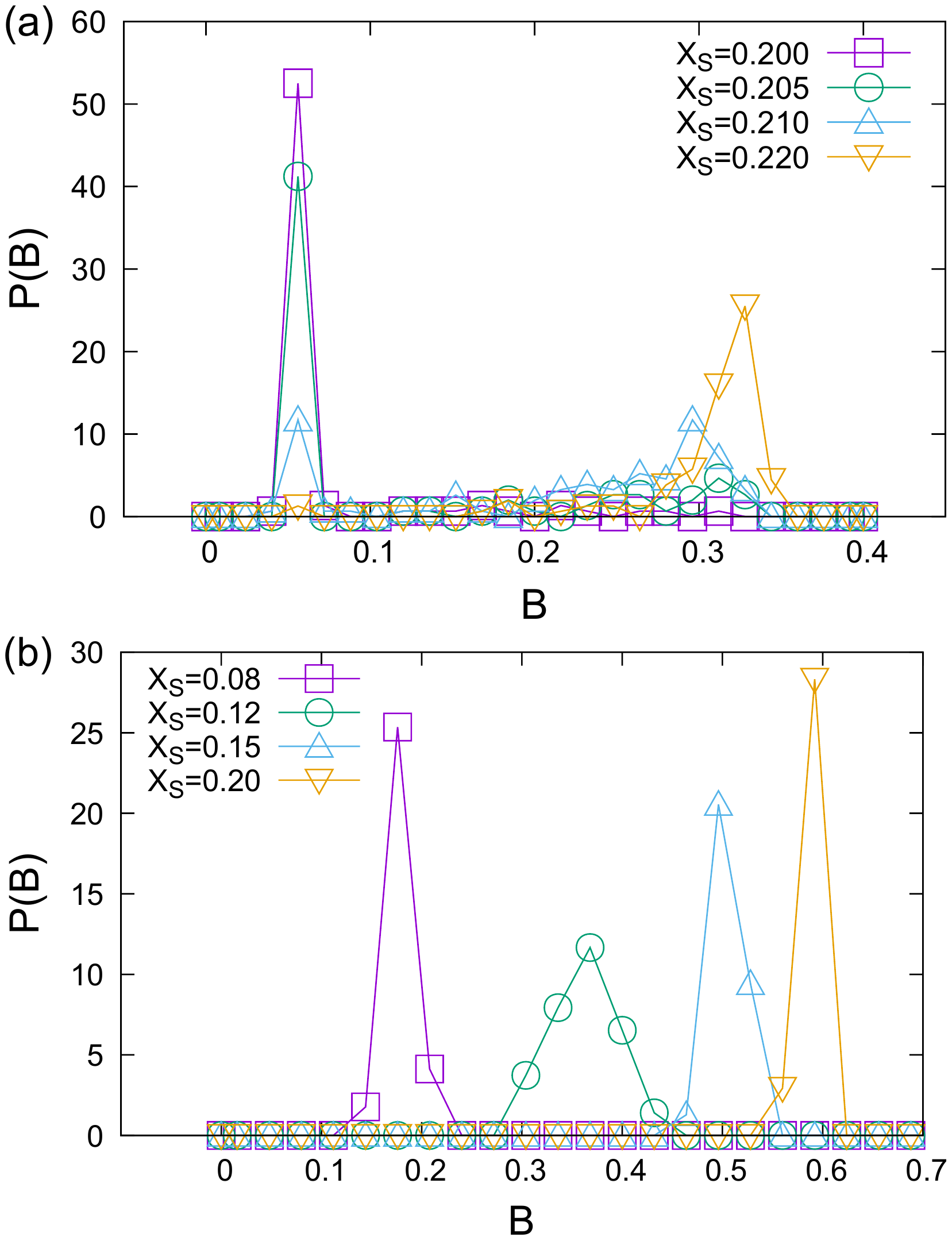}
\caption{
Distribution of bulk moduli $B$. At the pressure of the first-order transition, the bimodal distributions are observed.
(a) Distribution of $B$ at $P=10^{-3}$.
(b) Distribution of $B$ at $P=2 \times 10^{-2}$.
}
\label{fig:HISTBULK}
\end{figure}

\begin{figure}[t]
\includegraphics[width=\columnwidth]{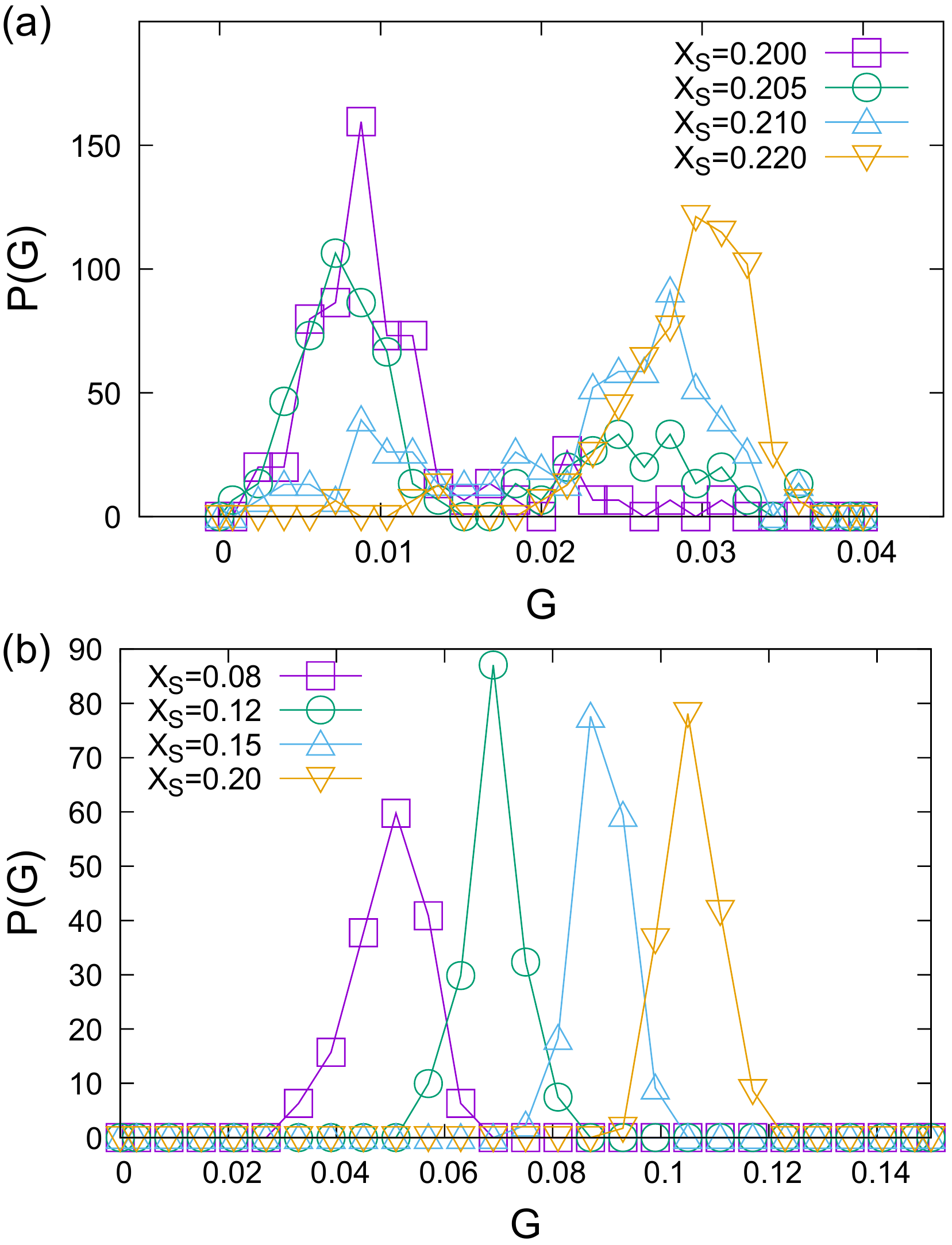}
\caption{
Distribution of shear moduli $G$. The bimodal distributions of the shear moduli become less apparent because of the fluctuation from the jamming transition.
(a) Distribution of $G$ at $P=10^{-3}$.
(b) Distribution of $G$ at $P=2 \times 10^{-2}$.
}
\label{fig:HISTSHEAR}
\end{figure}

Fig.~\ref{fig:Ave-mod} shows the values of the bulk and shear moduli averaged over different packing samples. 
Both moduli increase when $X_S$ increases. 
This result is consistent with the behaviors of $\langle R_S \rangle_{\rm sample}$ in Fig.~\ref{fig:AveRS}; the small particles participate in the rigid network and 
contribute to the rigidity of the system when $X_S$ increases. 
As in the case of $\langle R_S \rangle_{\rm sample}$, both moduli more dramatically increase when the pressure is decreased. 

Similar to the analysis in the previous sections, we calculate the probability distributions of the bulk and shear moduli over different packings samples.
Fig.~\ref{fig:HISTBULK} and Fig.~\ref{fig:HISTSHEAR} show the behaviors of the distributions of both moduli at high and low pressures.
At $P=10^{-3}$, the distributions of both moduli are bimodal as in the case of $R_S$.
The peak at smaller moduli corresponds to the L phase, and the other corresponds to the LS phase.
Thus, the elastic moduli also show discontinuous changes near the first-order transition in the jammed phase. 
Compare to the bulk moduli, the shear moduli have a broader distribution presumably because the fluctuation of the shear moduli becomes huge near jamming point $P \to 0$; the bimodal distributions in shear moduli are smeared by the critical fluctuations from the jamming critical phenomena.
At a higher pressure than the critical pressure, the bimodal distributions of both moduli are no longer observed. 
Therefore, the first-order phase transition is also characterized by the transition in mechanical properties.

\begin{figure}[t]
\includegraphics[width=\columnwidth]{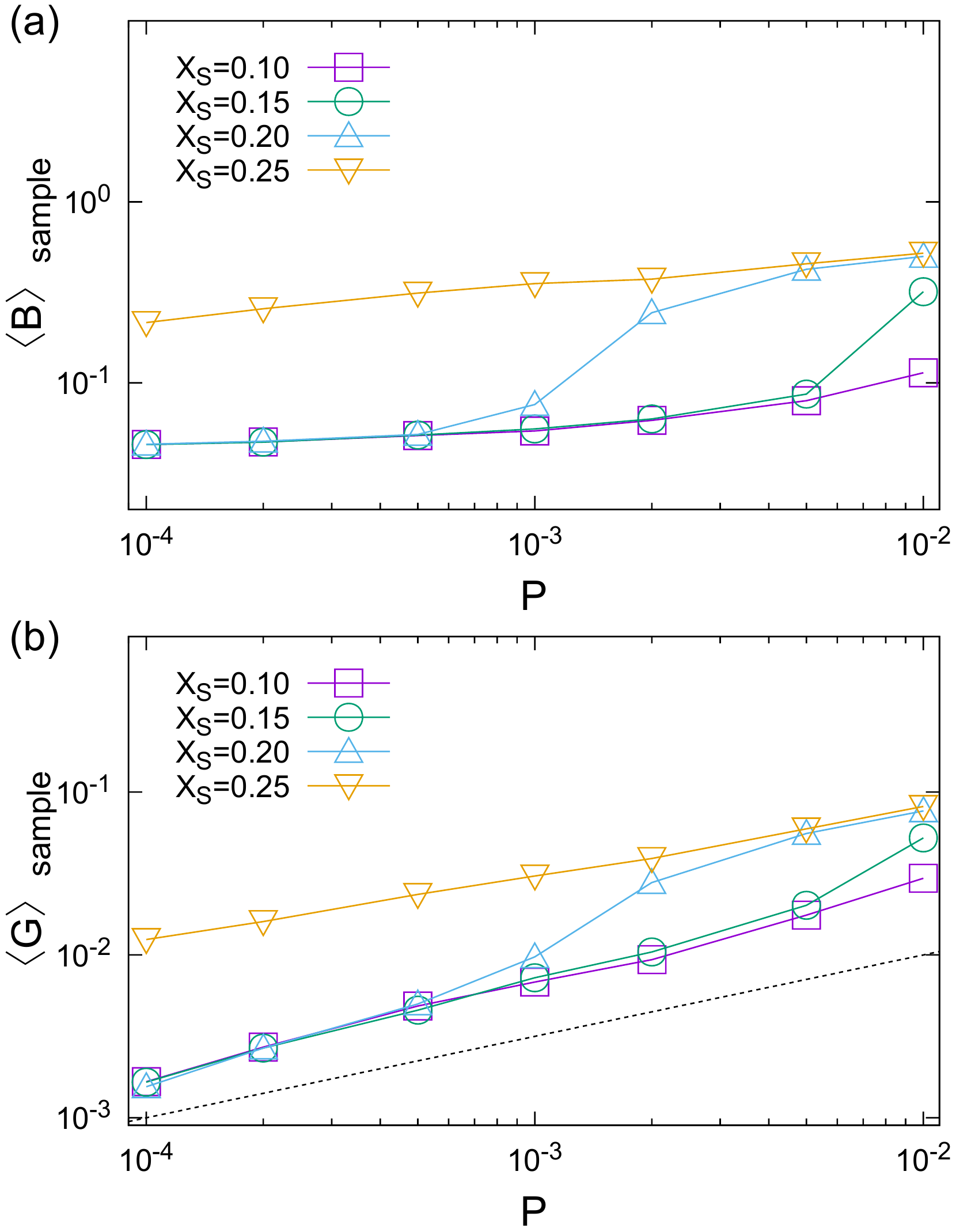}
\caption{
(a) Bulk modulus and (b) shear modulus plotted as functions of pressure $P$. 
The dotted line in (b) has a slope of $\frac{1}{2}$. 
}
\label{fig:BJSC}
\end{figure}

Finally, we discuss the pressure dependence of the moduli. 
In the monodisperse systems, many physical quantities are known to follow the critical power-law near the jamming transition.
It has been established that the bulk modulus is independent of the pressure of the system, while shear modulus depends on the square roots of the pressure.
However, it is not clear whether this jamming scaling is valid in the binary mixtures because of the discussed discontinuities of elastic moduli .
In Fig.~\ref{fig:BJSC}, we plot the average bulk modulus $\langle B \rangle_{\rm sample}$ and shear modulus $\langle G \rangle_{\rm sample}$ versus $P$ with fixed $X_S$.
Clearly, there are two branches in the plot: one branch is associated with small $X_S$, and the other is associated with large $X_S$. 
At intermediate $X_S$, we observe the jump from one branch to the other. 
In the lower branch, the systems are in the L phase, and the upper branch corresponds to the LS phase.
In each branch, bulk moduli $B$ are constant with pressure $P$, and shear moduli $G$ depend on the square root of pressure $P$
Hence, despite the discontinuities of the elastic moduli, the jamming scaling $G \propto p^{\frac{1}{2}}$ holds in each jammed phase.

\section{\label{conclusion}Conclusion}
In summary, we have numerically studied the structural and mechanical properties of the packings of binary mixture of particles with large size dispersity. 
We have presented strong evidences that the system exhibits the first-order phase transition between two distinct jammed phases: L phase, where only large particles are jammed, and LS phase, where both types of particles are jammed. 
This study was achieved by analyzing the statistics of the fraction $R_S$ of small particles that participate in the connecting network. 
The mean value of $R_S$ shows sharp increases at finite $X_S$, and the increase becomes progressively steeper when the system size increases. 
The probability distribution of $R_S$ exhibits typical behaviors of the first-order transition, and the susceptibility associated to $R_S$ linearly increases with the system size. 
We have also shown that the elastic moduli can be used as the order parameter of this transition. 
The LS phase is more rigid than the L phase. 
The probability distribution of the bulk and shear moduli behave similarly to those of $R_S$. 
Finally, we have shown that these moduli in each phase follow the critical power-law as in the monodisperse system. 


For nearly monodisperse systems, it has been established that the vibrational and transport properties exhibit the critical behavior near the jamming transition~\cite{Silbert2005,Wyart2005,Wyart2010,Vitelli_2010,DeGiuli2014,Mizuno2017,Shimada_2018,Mizuno2018}. 
As we established that the binary mixtures with large size dispersity exhibited another transition, it is interesting to study the vibrational properties of this system. 
We are now working in this direction. 

\begin{acknowledgments}
We thank K. Hukushima and H. Ikeda for useful discussions.  
We also thank B. P. Tighe for useful information on the result in two dimension 
This work was supported by JSPS KAKENHI Grants No. 
18H05225, 
19H01812,
19K14670, 
20H01868, 
20H00128.
The computations were partially performed using the Research Center for Computational Science, Okazaki, Japan.
\end{acknowledgments}

\appendix
\section{\label{CE}Formulation of elastic constants}
In this appendix, we introduce the linear response formalism to calculate the elastic constants.
We basically follow the formalism of Lemaitre \cite{Lemaitre2006}.
$U$ is a function of both particle positions ${\vec{r_i}}$ and strain tensor ${\eta}_{\alpha \beta}$.
The stress tensor $t_{\alpha \beta}$ is defined as
\begin{align}
t_{\alpha \beta} &= \frac{1}{V} \left( \frac{D U}{D \eta_{\alpha \beta}} \right)_{\eta_{\alpha \beta}=0}, \\ 
&= \frac{1}{V} \left( \frac{\partial U}{\partial \eta_{\alpha \beta}} + \sum_{i} \frac{D \vec{r_i}}{D \eta_{\alpha \beta}} \cdot \frac{\partial U}{\partial \vec{r_i}} \right)_{\eta_{\alpha \beta}=0},
\end{align}
where $\frac{D}{D \eta_{\alpha \beta}}$ is the derivatives that impose the mechanical equilibrium $\vec{f_i} = - \frac{\partial U}{\partial \vec{r_i}} = \vec{0}$, and the second term vanishes because of this condition.
The mechanical equilibrium leads to the following equation
\begin{equation}
\vec{0}=\frac{D}{D \eta_{\alpha \beta}}\left(\frac{\partial U}{\partial \vec{r}_i} \right)=\frac{\partial^2 U}{\partial \eta_{\alpha \beta} \partial \vec{r_i}} + \sum_{j} \frac{\partial ^2 U}{\partial \vec{r}_j \partial \vec{r}_i} \frac{D \vec{r}_j}{D \eta_{\alpha \beta}},
\end{equation}
where $ \frac{\partial ^2 U}{\partial \vec{r}_j \partial \vec{r}_i} = \bm{H}_{ij}$ is a $3 \times 3$ matrix with $ij$ components of dynamical matrix $\bm{H}$, and $ \frac{\partial^2 U}{\partial \eta_{\alpha \beta} \partial \vec{r_i}}=\vec{\Xi}_{i, \alpha \beta}$ is the non-affine force field.

Elastic constants are defined as the second-order derivatives of the energy by the strain tensor $\eta_{\alpha \beta}$ and have following form
\begin{equation}
C_{\alpha \beta \kappa \chi} = \frac{1}{V}\left( \frac{D^2 U}{D \eta_{\alpha \beta} D \eta_{\kappa \chi}} \right)_{\eta_{\alpha \beta}=0}.
\end{equation}
This is decomposed into two terms as follows
\begin{equation}
C_{\alpha \beta \kappa \chi} = \frac{1}{V} \left( \frac{\partial^2 U}{\partial \eta_{\alpha \beta} \partial \eta_{\kappa \chi}}
+ \sum_{i} \frac{D \vec{r}_{i}} {D \eta_{\alpha \beta}}
\cdot \frac{\partial^2 U}{\partial \vec{r_i} \partial \eta_{\kappa \chi}} \right)_{\eta_{\alpha \beta}=0}.
\end{equation}
Unlike the stress tensor, the second term is non-zero under mechanical equilibrium and is called the non-affine correction to the elastic constants.
The derivatives under constraint is replaced by the non-affine force field $\vec{\Xi}_{i, \alpha \beta} $, which leads to
\begin{equation}
C_{\alpha \beta \kappa \chi} = \frac{1}{V} \left( \frac{\partial^2 U}{\partial \eta_{\alpha \beta} \partial \eta_{\kappa \chi}}
- \sum_{ij} \vec{\Xi}_{i, \alpha \beta} \bm{H}_{ij}^{-1} \vec{\Xi}_{j, \kappa \chi} \right)_{\eta_{\alpha \beta}=0}.
\end{equation}


The second term is rewritten by diagonalizing dynamical matrix $\bm{H}$
\begin{equation}
\sum_{ij} \vec{\Xi}_{i, \alpha \beta} \bm{H}_{ij}^{-1} \vec{\Xi}_{j, \kappa \chi} = \sum_{n} \frac{\Xi_{\alpha \beta}^n \Xi_{\kappa \chi}^n}{\lambda^n}.
\end{equation}
Here, the nth eigen values and eigen vectors of $\bm{H}$ are represented as $\lambda^n$ and $\vec{\Psi}^n$, and the inner product of $\vec{\Xi}_{\alpha \beta}$ and $\vec{\Psi}^n$ is $\Xi_{\alpha \beta}^n$.
\begin{equation}
C_{\alpha \beta \kappa \chi} = \frac{1}{V} \left( \frac{\partial^2 U}{\partial \eta_{\alpha \beta} \partial \eta_{\kappa \chi}}
- \sum_{n} \frac{\Xi_{\alpha \beta}^n \Xi_{\kappa \chi}^n}{\lambda^n} \right).
\end{equation}

By evaluating elastic constants $C_{\alpha \beta \kappa \chi}$, bulk moduli $B$ and shear moduli $G$ are calculated as follows, 
\begin{equation}
B = \frac{1}{9} \left( C_{xxxx}+C_{yyyy} + C_{zzzz} + 2 C_{xxyy} + 2 C_{yyzz} + 2 C_{zzyy} \right),
\end{equation}
\begin{equation}
G = \frac{1}{3} \left( C_{xyxy}+C_{yzyz} + C_{zxzx} \right).
\end{equation}



\bibliography{binary}
\bibliographystyle{unsrt} 


\end{document}